# Rhythm Zone Theory: Speech Rhythms are Physical after all (V09)

*Dafydd Gibbon and Xuewei Lin, Jinan University, Guangzhou, P.R. China*

## Abstract

Speech rhythms have been dealt with in three main ways: from the introspective analyses of rhythm as a correlate of syllable and foot timing in linguistics and applied linguistics, through analyses of durations of segments of utterances associated with consonantal and vocalic properties, syllables, feet and words, to models of rhythms in speech production and perception as physical oscillations. The present study avoids introspection and human-filtered annotation methods and extends the signal processing paradigm of amplitude envelope spectrum analysis by adding an additional analytic step of edge detection, and postulating the co-existence of multiple speech rhythms in rhythm zones marked by identifiable edges (Rhythm Zone Theory, RZT). An exploratory investigation of the utility of RZT is conducted, suggesting that native and non-native readings of the same text are distinct 'sub-genres' of read speech: a reading by a US native speaker and non-native readings by relatively low-performing Cantonese adult learners of English. The study concludes by noting that with the methods used, RZT can distinguish between the speech rhythms of well-defined sub-genres of native speaker reading vs. non-native learner reading, but needs further refinement in order to be applied to the paradoxically more complex speech of low-performing language learners, whose speech rhythms are co-determined by non-fluency and disfluency factors in addition to well-known linguistic factors of grammar, vocabulary and discourse constraints.

## 1 Rhythm in a semiotic framework

Rhythms in music and language are semiotic events: regularly repeated structured temporal patterns of human experience in performing and perceiving music, dance and speech, and, more metaphorically, to events of non-human origin such as animal sounds, and to regularly repeated spatial patterns in the visual arts and in the dynamics of natural phenomena. The aim of the present study is to examine the frequencies of speech rhythms in the newly developed framework of Rhythm Zone Theory (RZT), and to illustrate a possible application domain in the field of foreign language fluency assessment, using two non-fluency markers derived from RZT. These two aspects of theory and practice relate to two of the main research interests *amicae optimae laudataeque libri huius*. This work is exploratory and concerned with methodological issues; the case study is illustrative of the method rather than primarily evidential. The background to the work is formulated in Time Type Theory (Gibbon 1994, 2006; Gibbon & Griffiths 2017), which provides an ontology of four linguistically relevant time concepts: (1) *abstract categorial time*, as in duration contrasts between long and short vowels; (2) *abstract relational ('rubber') time*, as in the sequential and hierarchical relations postulated in linguistic descriptions; (3) *clock time*, as in measurements of time points and intervals in a speech signal and (4) *cloud time*, as in the intuitively perceived timing of actual utterances as they are made. The present study is concerned with clock time.

From a semiotic point of view, rhythms have functional, formal and physical characteristics: functions in communication, forms as sequences, hierarchies and parallel streams, and physical characteristics in the movements of a musician or in the movements of the organs of speech in the vocal tract. In the long history of the scientific treatment of rhythms, the complex interactions of the three semiotic principles of function, form and physics have often, intuitively, been taken to reflect emotions associated with rhythmical aspects of human behaviour and perception, such as faster and slower breathing, heartbeats, or limb movements which are determined by the properties of human anatomy and physiology. The literature on these topics is legion, and the present study focuses only on a very small part of this literature.

The functional aspects of rhythms are perhaps the most complex and the least researched: the importance of rhythms as cohesive means of framing speech and music into coherent, manageable information patterns is perhaps most obvious in speech, and the emotional heartbeats of rhythm are perhaps most obvious in music, but rhythms in both speech and music share cohesive and emotional functions. A more general functionality of rhythm is described in a thought experiment of whether there could be a world with time as its only dimension (Strawson 1959), in which dynamic changes in amplitude (and thus also rhythms) may be interpreted as approaching and disappearing sound source objects.

The formal characteristics of rhythms are linear and hierarchical patterns of sounds in time. In linguistics, particularly in the phonology of sentences and words, there is an extensive field of research in modelling these patterns, most clearly represented in the *nuclear stress rule* and the *compound stress rule* of generative phonology (after Chomsky & Halle 1968), the *metrical grid* (after Liberman & Prince 1976), the *prosodic hierarchy* (Selkirk 1984), *beats and binding* (Dziubalska-Kołaczyk 2002) and other variants of and successors to generative phonology. In linguistic studies a certain scepticism about phonetic studies of rhythm reigns, suggesting that rhythms are primarily cognitive constructs, or even not identifiable in physical terms at all.

The physical characteristics of rhythms are found in the dynamics of the production of sounds with musical instruments and the voice, and in the perception of these sounds. In musicology and in phonetics, there have been many approaches to capturing, describing and explaining the physical characteristics of rhythm. In phonetics, much effort has been spent on the investigating 'the' rhythm of a language, dialect or idiolect with various phonetic methods, for instance by investigating repeated temporal patterns aligned with syllables and words. These studies have not been particularly successful, and many have relied on human filtering of the speech signal through the procedures of manual annotation (and automatic annotation, i.e. annotation by supervised machine learning involving bootstrapping with manual annotations). More success has been achieved by studies of rhythms as oscillations (see Section 3).

A terminological clarification is necessary at this point. Speech involves approximate frequency ranges of three different types, of which only the first is relevant for the present study:

1. From 0 Hz ... 20 Hz: the domain of the frequencies and their phases which characterise the rhythms of speech sounds, syllables, words, phrases and larger discourse units, which determine the low frequency outline (the *amplitude envelope*) of the speech signal; events in this frequency range are perceived as separate beats rather than as tones.
2. 80 Hz ... 400 Hz (adult male and female voices): the domain of the *fundamental frequency* of the voice, which relates to tones, pitch accents and intonation, the domain usually shown in F0 tracks, 'pitch' tracks; in this frequency range, events are perceived as tones.
3. 80 Hz ... 4000 Hz: the domain of the *spectral formants* shaped by the oral and nasal cavities of the vocal tract, which characterise vowels and consonants and voice quality, the domain usually shown in spectrograms; particularly in the mid and upper sections of this frequency range, events are perceived as sound qualities.

The following sections concentrate on these temporal physical characteristics, and show that 'the' rhythm of a language is best not thought of as 'the' rhythm at all: there are many rhythms, in different temporal domains, and each of the rhythms is highly variable both in frequency and in phase. The physical characteristics of speech rhythms are measurable and visualisable using signal processing methods, also in neurophysiological domains. Finally, an application of this recent methodology in a practical field will be demonstrated: the capturing of temporal non-fluency in readings by low proficiency adult Cantonese learners of L2 English.

## 2 Irregularity and isochrony: the annotation method

One basic method of investigating speech timing is by aligning linguistic units with segments of the speech signal, measuring the duration of these units, and performing descriptive statistical analyses and structure building on these duration measurements. Annotation (also known as labelling or phonetic alignment) is a method of pairing the components of a transcription of a speech recording (*labels*) with *time-stamps* which indicate the beginning and end, or the beginning and length (more rarely: the middle) of these components, with the aid of speech visualisation software. For manual annotation, the most popular software tool is *Praat* (Boersma 2001); cf. also *Wavesurfer* (Beskow and Sjölander 2004), *Transcriber* (Barras et al. 2001), *Annotation Pro* (Klessa and Gibbon 2014). The annotations produced with each tool are largely interconvertible. For semi-automatic annotation, a convenient software tool is *SPPAS* (Bigi 2015). The first tools for the annotation method were originally developed in speech technology, for bootstrapping supervised machine learning procedures in statistical automatic speech recognition. Three main kinds of approach based on the annotation method have emerged: one-dimensional, two-dimensional and three-dimensional irregularity and isochrony models.

The one-dimensional approaches are based on the calculation of an index based on the descriptive statistics of label durations. These approaches have been used to investigate the temporal typology of different spoken languages, and are necessarily based on an assumption that the kind of unit to be labelled (vocalic and consonantal sequences; syllables; feet; word) is universally found in all languages. The simplest index is the *standard deviation* of the durations of the units of the relevant type (for variants of this index cf. Roach 1982, Scott et al. 1985). One problem with this approach is the hidden factor of *speech tempo*: the unit rate per second may vary for different reasons during an utterance. This hidden factor of speech tempo variation is abstracted out by the *normalised Pairwise Variability Index*, which averages the normalised duration differences between neighbouring units, yielding an index with an asymptote of 200: $0 \leq i < 200$. The lower the index, the more regular the timing of the units measured. The original version of the *nPVI* formula is recast here in order to express the functionality of the operations more transparently:

$$nPVI = 100 \times mean(absolute(d_i - d_{i+1}) / mean(d_i, d_{i+1}))$$

Formally, the *nPVI* is a variant of the Canberra Distance measure, applied to a pair consisting of a vector and the identical vector shifted to the right (or left) by one position. Canberra Distance is in turn a normalisation of the well-known Manhattan Distance measure (Taxicab Distance). Typical *nPVI* values are discussed by Low et al. (2000), Grabe et al. (2002) and many others. Values related to the present study are discussed by Gibbon & Yu (2015): with measurements based on read-aloud texts, Chinese, said to be a *syllable-timed* language, has *nPVI* values around 35, Farsi around 45 and English, as a *stress-timed* or *foot-timed* language, around 60.

The one-dimensional indices are measures of irregularity or relative isochrony, and provide a useful heuristic for the initial study of low frequency speech timing, but the popular term 'rhythm metrics' rather than 'isochrony metrics' is seriously misleading: the rhythmic criterion of similarity of repeated utterance segments, the alternation or 'boom-de-boom boom' component[1] of rhythm, is actually factored away by taking the absolute value of the subtraction. The subtraction operation reduces timing relations to a local binary isochrony relation. Nolan et al. (2014) disputed the point that the *nPVI* defines a binary relation, but the binarity of the modified Canberra Distance measure can hardly be denied. Speech rhythms are not necessarily binary trochaic and iambic patterns, however, which the binary subtraction operation implies, but can be ternary anapaestic (*weak-weak-strong*), dactylic (*strong-weak-weak*), amphibrachic (*weak-strong-weak*), or even more complex. These more complex patterns are not captured by the currently proposed isochrony indices; if desired, however, arbitrarily complex *n*-ary relations can easily be defined by using a vector shift of more than one position. It would also be helpful to note the standard deviation of the differences in order to estimate the validity of the index, since the index varies wildly throughout utterances, but this is not done in the available literature. A two-dimensional version of the *PVI* approach was developed by Nolan and Asu (2009) and Asu and Nolan (2009), combining analyses of foot patterns (cf. also Roach 1982) and syllable patterns.

A different two-dimensional model was developed by Wagner (2007), in which rhythm is represented as a 'neighbour relation' (cf. Figure 1): durations are normalised by z-score transformation, and durations of neighbouring pairs are represented on the *x* and *y* dimensions of a scatter plot. If the unit type is syllable, then 'syllable timing' is indicated if the pairs are distributed evenly and symmetrically around the zero points of the two axes, while 'foot timing' is indicated if the distribution is highly skewed to the upper right quadrant (above the means), with the majority of points being in the negative (below the means) lower left quadrant of the scatter plot. The criticism of restriction to binary relations still applies, but only partially: the skewed distribution of points between the negative and the positive quadrant may indicate that the timing pattern is non-binary.

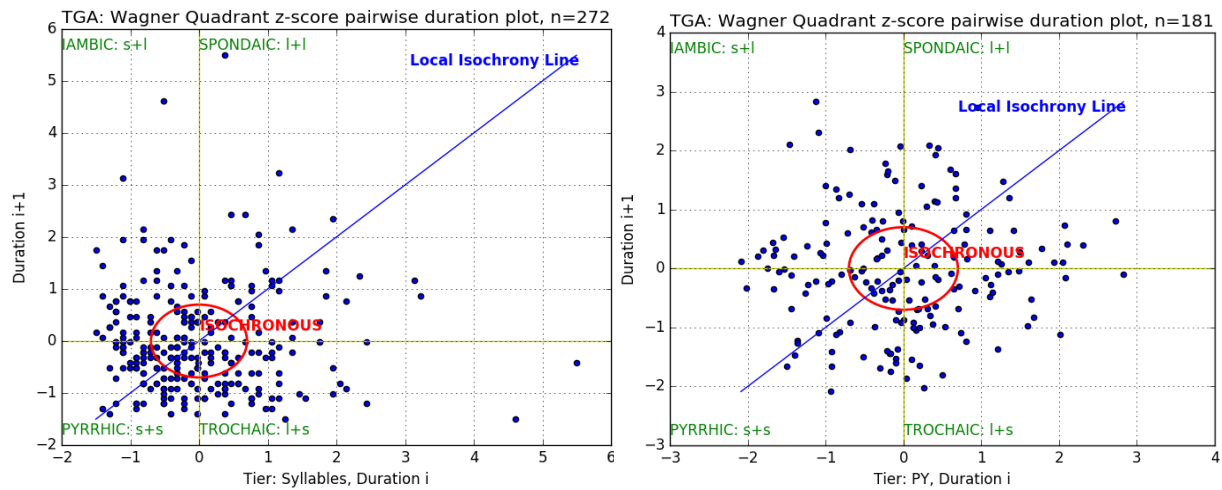

*Figure 1: Wagner quadrant representation of the four binary duration relations between adjacent syllables, relative to mean duration (z-score transformed): English (left); right skew, clustering in bottom left quadrant, Mandarin (right, no skew, clustering around the mean).*

A three-dimensional model (cf. Figure 2) was developed by Gibbon (2003), also based on durations of annotated categories such as words, with duration relations between neighbours being used to hierarchically adjoin units in a tree structure (the first and second dimensions), with the option of strict binary or non-binary adjunction, and the option of *strong-weak* or *weak-strong* pairing, with *weak* and *strong* interpreted as *shorter* and *longer* respectively (the third dimension). The resulting tree patterns are mapped to grammatical and discourse structures.

---

1  Informal remark by Fred Cummins, lecture ca. 2003.

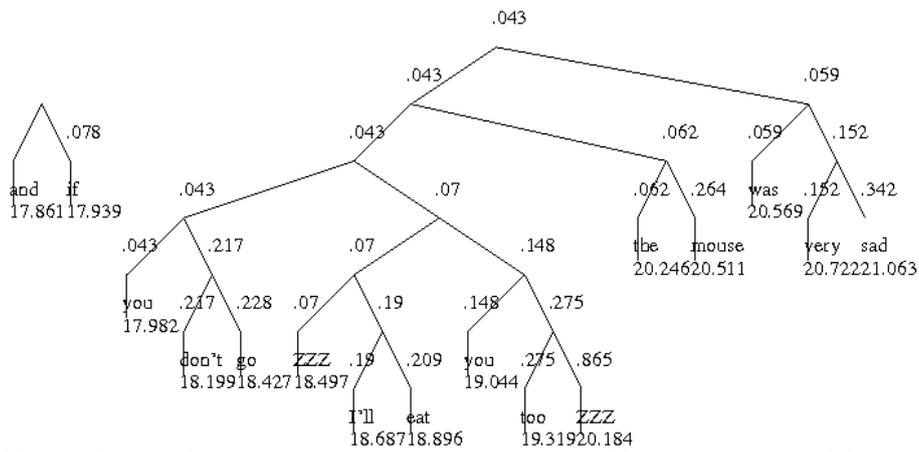
*Figure 2: Time Tree spanning 3 seconds of a story* (*The Tiger and the Mouse*), *iambic time relations.*

To summarise, the annotation oriented approaches are useful transforms of the speech signal and the isochrony metrics have the indisputable merit of being easy to calculate, but they are neither rhythm models nor rhythmicity indices. Their epistemological status is that of heuristic strategies which are human-filtered by manual annotation (or semi-automatic annotation, statistically trained on prior manual annotations in a supervised machine learning procedure). They are in need of a theoretical foundation.

## 3 Rhythm frequencies and Rhythm Zone Theory

The present approach seeks to provide an objective measure of rhythm which is independent of human-filtered annotations and derives timing properties directly from the speech signal. The basic idea is that speech is is a complex signal consisting of a carrier signal which is simultaneously modified in two ways: by *frequency modulation* (functioning as tone, pitch accent, intonation) and by *amplitude modulation* (functioning as phones, syllables, words), and that these two types of modulation both have rhythmic properties. This approach is represented from the point of view of speech production by the oscillator theories of rhythm (e.g. Cummins and Port 1998, O'Dell and Nieminen 1999, Barbosa 2002, Inden et al. 2012).

From the point of view of speech analysis (and speech perception modelling), the signal is decomposed by a *frequency demodulation* function (in phonetic terminology: F0 extraction, 'pitch' tracking) in the frequency range 80...400 Hz, and by a very complex *amplitude demodulation function*, which maps the signal in the range 80...4000 Hz to categorial units of phones, syllables, words, etc. The amplitude demodulation function also tracks the outline of the waveform (the amplitude envelope) in the range 0...20 Hz, the low frequencies which underlie perception of the long-term rhythmic patterns of phones, syllables, words, etc. These low frequency rhythmic patterns and their phases are revealed by spectral analysis of the amplitude envelope. A paradigm has developed in which the amplitude envelope demodulation method of rhythm modelling has been proposed as a more adequate approach to the modelling of rhythm than the isochrony metric approach (Todd 1994, Cummins et al. 1999, Tilsen and Johnson 2008, Hermansky 2010, Leong et al. 2014, Leong and Goswami 2015, Ludusan et al. 2011, Tilsen and Arvaniti 2013, Tilsen 2016, He and Dellwo 2016, Varnet et al. 2017, Ojeda et al. 2017, Gibbon 2018).

Three steps are primarily involved, with a fourth step for spectral edge detection (rhythm zone boundary detection) introduced recently (Gibbon 2018). The four steps are illustrated in Figure 3:
1. Input the signal from a recording (instruction here: "Count as quickly as possible from one to thirty!") and store the *waveform*, illustrated as an oscillogram (Figure 3, top left). The signal length is about 10 seconds, the thirty words occur with frequency of about 3 per second, and the periodicity corresponds to an average word length of about 0.3 seconds.

2. Extract the *amplitude envelope* of the waveform. To do this, in formal terms a *Hilbert Transform* is performed on the waveform segment of interest, and the absolute value of the transformed signal is low-pass filtered, leaving the *positive amplitude envelope*, the outline of the shape of the signal (Figure 3, bottom left). In the implementation used here, instead of the Hilbert Transform an efficient peak-tracing algorithm was used on the rectified (absolute) signal, also yielding the positive amplitude envelope. By extraction of the envelope, frequencies in the range 0...20 Hz are selected, and higher frequencies discarded.

3. Extract the frequency spectrum of the amplitude envelope within the range 0...20 Hz (Figure 3 top right) by means of a *Fast Fourier Transform* (*FFT*) on the amplitude envelope in order to extract the low frequency spectrum of the amplitude envelope, the *Amplitude Envelope Spectrum* (AES). The AES contains the low frequencies with which phones, syllables, words, etc. occur. The phases of these frequencies are not focussed in this study. The frequencies and their phases are fairly regular in the present counting data type, which is reflected in rather clear and plausible prominent frequencies.

4. Conceptualise different low frequency spectral ranges below 20 Hz as *rhythm zones* (RZ) and perform *edge detection* on the AES (introduced to spectrum processing by Jassem et al. 1983) to identify the rhythm zone edges (RZE). At the present stage, an elementary variety of edge detection by differencing the digital envelope spectrum is used; other algorithms are under development. The result of edge detection is the *Amplitude Envelope Difference Spectrum* (AEDS), shown in Figure 3, bottom right.

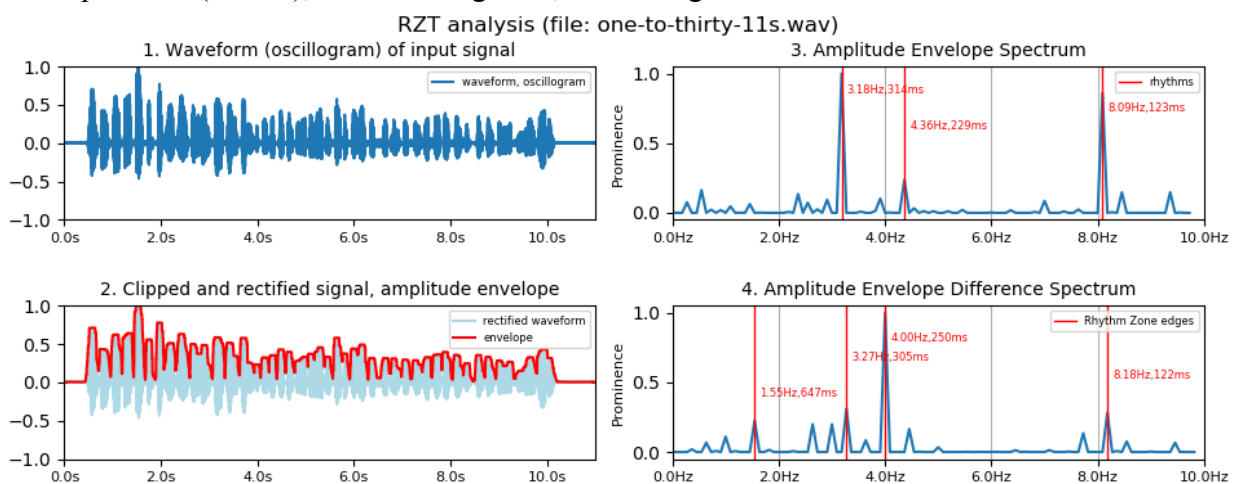

*Figure 3: Counting from 1 to 30 (English). 1. Top left: Waveform (oscillogram). 2. Bottom left: Waveform, showing rectified waveform and amplitude envelope modulation. (3) amplitude envelope spectrum. (4) amplitude envelope difference spectrum.*

Having defined the signal processing procedure, a technical definition of 'rhythm' can be given: a *rhythm* is defined as a prominent frequency in the amplitude envelope spectrum. A series of rhythms defined in this way is referred to as a *rhythm spectrum*. Frequency ranges containing a rhythm in this sense are referred to as a *rhythm zone*. The top right graph in Figure 3 shows two prominent frequencies in the amplitude envelope spectrum. As predicted from the instruction to count from one to thirty as quickly as possible, and from a rough inspection of the oscillogram, the AES shows a rhythm at 3.18 Hz, indicating a periodicity of 314 ms (corresponding approximately to syllable articulation rate). The AEDS displays the edges of the main rhythm zones. Accordingly, the present approach is referred to as *Rhythm Zone Theory* (RZT).

The RZT algorithm is entirely agnostic about the linguistic units involved, but in the present example, taking the a priori instruction to the speaker to count from one to thirty into account, the main linguistic unit can be identified as a word, either a monosyllable (e.g. *one*, *two*, etc.) or polysyllabic (e.g. *seven*, *eleven*, … *thirty* etc.). There are minor rhythms, the strongest of which has a rhythm at 8.09 Hz, with a period (unit length) of 123 ms, for which the relevant linguistic unit can be tentatively identified as the syllable components of polysyllabic words. Higher frequency rhythms are associated with weak and reduced syllables, with syllabic sonorants, and with the rhythms of syllable constituents.

Figure 3 visualises a clear case of rhythm identification based on RZT: counting is a highly regular genre of speech production. In contrast, Figure 4 visualises the reading of a fable, a genre which is open to more rhythmic variation and to potential dominance of the spectrum by higher ranking discourse rhythm types, rather than by the syllable and word domain rhythms of Figure 3. The most prominent rhythm shown in Figure 4 (top right) is 0.18 Hz (5.596 s) corresponds to about half of the recording, determined by the long pause shown in the oscillogram at approximately 6.5 s. The rhythms at 1.36 Hz (733 ms) and 1.82 Hz (549 ms) indicate a dominance of phrasal and discourse rhythms over syllabic, foot or word rhythms. Other less prominent spectral frequency peaks are associated with shorter components of the speech signal, for example at about 3.5 Hz (286 ms), for strong or stressed syllables and at about 5.5 Hz (181 ms) for weak syllables and longer phones. In the more complex narrative data of Figure 4, the shifting phases of the frequencies, which are not separately captured in this study, lead to a more complex and 'noisy' spectrum visualisation.

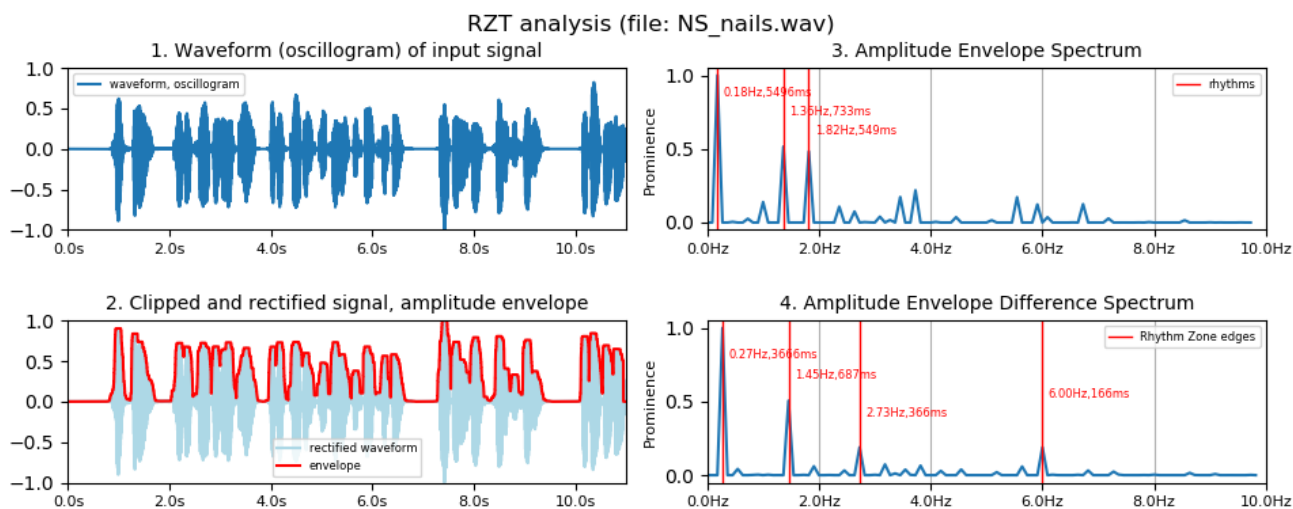

*Figure 4: Beginning of a story (NS English). The prominence of rhythms and rhythm zone edges is shown by relative height of the peaks in the spectra.*

Summarising: Rhythm Zone Theory (RZT), the theoretical foundation of the present four-step approach to phonetic rhythm analysis, makes the following explicit claims:

1. RZT has no need of subjectively influenced transcription and annotation, but applies directly and reproducibly to the speech signal.
2. RZT postulates that speech has rhythms of many different frequencies, putatively correlating with phone, syllable, foot and word timing patterns through phrase and discourse timing patterns, and that rhythms may vary in frequency and phase through longer discourses.
3. RZT conceptualises rhythms dynamically in terms of frequencies in Hertz, rather than statically in terms of the durations of pre-defined linguistic units in seconds or milliseconds, enabling a rhythm to be defined precisely and completely as a frequency in the long-term amplitude envelope spectrum associated with temporal phase shifts. Clearly, durations and rhythms have a simple relation, but the mindset underlying a frequency model differs greatly from the mindset underlying the descriptive statistics of unit durations.
4. RZT defines rhythms as occurring in rhythm zones, that is, the frequency ranges which surround rhythms; rhythm zone boundaries are fuzzy, partly caused by shifting oscillation phases during utterances.
5. RZT provides empirical signal-based evidential grounding which can be utilised in *post hoc* validation of how linguistic units align with physical rhythms, for example using annotations and annotation-mining techniques (Gibbon and Yu 2015). The rhythm alignments range from speech sound rhythms through syllable, word and phrase rhythms to discourse rhythms.

## 4 The use case of foreign language 'fluency' assessment

### 4.1 The background

The assessment of proficiency in a foreign language is as complex as the language itself, with the degree of proficiency understood as distance from native-like performance. In the present context, we understand fluency in a narrow sense, to refer primarily to prosodic aspects of proficiency in spoken language. We distinguish between four main kinds of fluency in this area and characterise the polysemy of 'fluency' through proof by contradiction in terms of four antonyms: *non-fluency* (due to lack of knowledge of a language); *disfluency* (due to lack of practice in a language); *impediment* (such as stuttering) and *aphasia*. The boundaries between the categories may be fluid. For the area of foreign language learning and teaching, the relevant category is non-fluency. Disfluency plays an important role, but it may also be a characteristic of native speakers. Impediment or aphasia may also play a role in some cases. Almost paradoxically, non-native speech is more complex than native speech, at least in relation to non-fluency, though not in terms of grammar, vocabulary and discourse strategies.

Within the area of non-fluency based on lack of knowledge, there are many potential *non-fluency markers*, from the discourse rank through phrasal and word rank to speech sounds. Among these are the prosodic non-fluency markers of rhythm and melody, which occur at every rank. The present contribution excludes melodic non-fluency marking of intonation, stress accent, tone and pitch accent and concentrates on selected aspects of speech timing.

Many timing related fluency or non-fluency markers have been discussed in the literature, including speech rate, phonation time ratio (speech:pause ratio), syllable pruning, mean length of interpausal unit ('length of run'), ratio of silent pauses, ratio of filled pauses, number of pauses, total pause time. Timing measures which explicitly involve linguistically identified categories, such as mean words per minute, self-repairs, repetitions, reformulations, replacements, false starts, hesitations, are excluded from the present study. Partial overviews are provided by Ellis (2009), Kuhn and Stahl (2000), Lambert and Kormos (2014). Other markers based on the isochrony metrics discussed in Section 2 have been used in the context of foreign language learning and teaching, most comprehensively by White and Mattys (2007) and White et al. 2012). The majority of these markers are identified via manual annotation techniques of various kinds, including otherwise objective methods which use automatic speech recognition, bootstrapped by the human-filtering training step of prior manual annotation (Cucchiarini et al. 2000, 2002; Zechner et al. 2009).

The present exploratory case study has a somewhat different aim: to outline the potential of the RZT approach for characterising learner speech in terms of rhythms. The main assumptions are:

1. Native Speaker (NS) reading and Non-Native Speaker (NNS) reading are different specialised 'sub-genres' of the general genre of read speech, even when NNS speech is an imitation of NS speech rather than just reading: a non-native learner reading English or Chinese is not dealing with exactly the same genre as a native speaker reading English or Chinese, because of non-fluency issues.
2. The envelope spectral properties of reading in the NNS sub-genre are more similar to each other than to those of the NS reading sub-genre, and can be distinguished automatically from NS reading on the basis of Rhythm Zone Theory.

**4.2 Method**

The participants are 12 EFL students selected from two college English courses at Jinan University (JNU), Guangzhou, China, with a relatively low proficiency level. The courses are designed for non-English major or minor students from various departments, and span three semesters (College English Elementary, Intermediate and Advanced, from September 2017 to December 2018). The classes are facilitated by a mobile app named *Moso Teach*,[2] which enables the students to record their speech in and after class.

Data are collected from recordings of the same text, an exhortative fable, by the 12 students in 2017 and 2018, totalling 24 recordings. In addition, a recording of the same text by a native speaker was taken from a JNU in-house EFL course and used for comparison.

A number of practical contributory factors to the heterogeneity of the data must be taken into account. The recordings were deliberately made in a realistic scenario: a classroom situation without direct supervision of the recording process, in a relatively noisy environment (cf. the high noise floor shown in the first second of the oscillogramm of Figure 6), with different recording devices (smartphone brands) with different automatic gain control properties and coding standards, with different overall speech loudness due to distance from microphone, and therefore different signal-to-noise ratios. The software implementation was designed to overcome these practical hurdles as far as possible using different kinds of filtering (Figure 5).

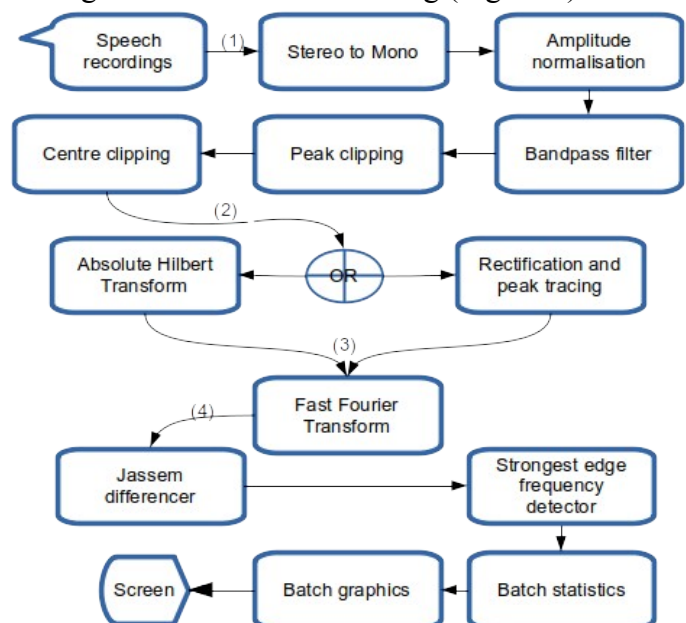

*Figure 5: Data flow in the RZT implementation with numbering showing the four steps of the RZT procedure.*

---

2   https://www.mosoteach.cn

An RZT analyser with a command line interface and graphic output was implemented, operationalising the four steps of the RZT methodology (Figure 5). Versions of the Python 2.7 code are available on GitHub and are interoperable on Linux, Windows and MacOS. The 24 recordings were automatically analysed and the following properties extracted:
1. The first 6 most prominent frequencies in the AES, and in the AEDS.
2. Standard deviation based features: (1) SD of the selected spectrum region; (2) SD of the spectrum region lower than the most prominent frequency; (3) SD of the spectrum region higher than the most prominent frequency; (4) mean difference between (2) and (3); (5) mean ratio between (2) and (3).

**4.3 Results**

**4.3.1 Visualisation**

Figure 3 and Figure 4 show relatively 'well-behaved' speech: regular counting, and story reading by a native speaker. The NNS story readings introduce a whole new range of noisy non-fluency factors on speech timing, which are predicted to be detectable by RZT analysis. In principle, it should also be possible to distinguish between different degrees of NNS-ness by looking for rhythms in the different rhythm zones determined by the many phrasal and discoursal factors which affect the timing of reading performance. In contrast to the relatively regular patterns of Figure 3 and Figure 4, Figure 6 shows same story as in Figure 4 being read by an adult Cantonese learner of English. The NNS patterns are rather different from those of the NS, in this case clustering around the 5 Hz, 200 ms periodicity area, implying that the major rhythm is syllabic. Auditory inspection of the recording shows that the reading is very mechanical strong syllable timing, as in the Cantonese source language or due to a handling strategy for non-fluency.

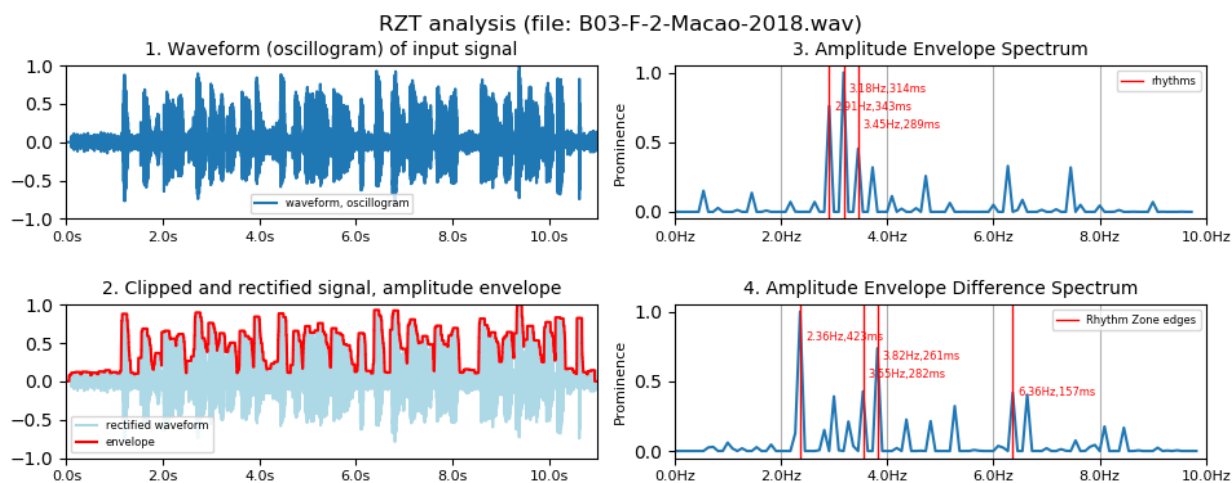

*Figure 6: Beginning of a story (NNS Cantonese English). The prominence of rhythms and rhythm zone edges is shown by relative height of the peaks in the spectra.*

**4.3.2 Rhythms as prominent frequencies**

For the first exploratory investigation, the first 60 s segment of each recording was analysed and the 6 most prominent frequencies were extracted as basic rhythms. Clearly, for a full and quantitatively interesting experiment far more data are needed in order to confirm the claims, but the aim here is to use visualisations to probe for basic trends when RZT is applied, and to open up a path to future research. No definitive claim to statistical significance is made at the present stage. The means for the 6 prominent frequencies were plotted for the 12 NNS against the 6 NS prominent frequencies, for 2017 and 2018 (Figure 7).

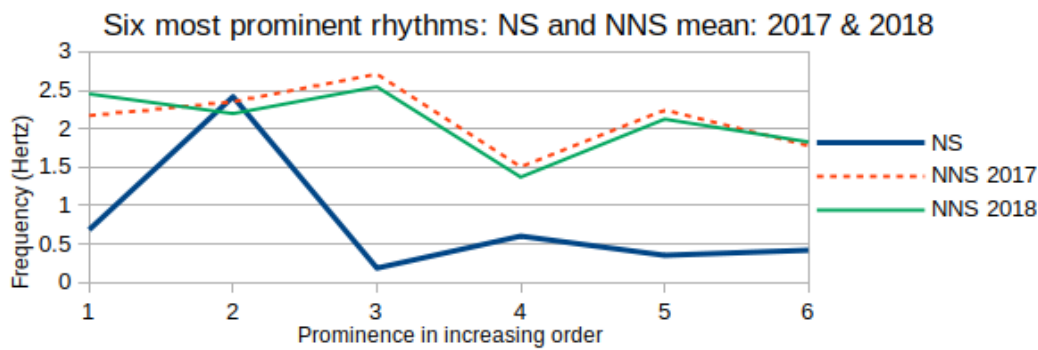

*Figure 7: Prominent rhythms: means for NNS against NS values for each prominence position.*

The first conspicuous feature of the visualisation is the distance between the NS rhythms and the NNS rhythms: Pearson's $r = 0.04$, Gosset's two-tailed paired t-test shows a significant difference, $\alpha = 0.05$ and the mean difference is 1.3 Hz. The second conspicuous feature is the very close similarity of the NNS group in 2017 and 2018, with a barely perceptible lower trend for 2018; Pearson's $r = 0.97$, Gosset's t-test shows no significant difference, and the mean difference is indeed only 0.07 Hz. The third conspicuous feature is the close clustering of the NNS rhythms in the rhythm zone between about 1.8 Hz and 2.7 Hz (SD = 0.14 Hz), corresponding to periodicities of 0.6 s and 0.4 s respectively. Identification of the linguistic units involved, presumably syllables or words, would involve detailed analysis of an annotation of the signal, but that is beyond the signal processing remit of the present study.

The conclusions to be drawn tentatively from the relations visualised in Figure 7 are (1) that RZT does not distinguish between different stages of low-performing learning; (2) RZT distinguishes clearly between low-performing (NNS) and high-performing (NS) categories; (3) in the low performing category (NNS) the faster word or long syllable rhythms dominate, while in the high performing (NS) category slower phrasal rhythms dominate.

### 4.3.3 Variability of rhythm zones

For the exploratory analysis of RZ variability, the first 170 seconds of each recording are divided into 16 consecutive segments of 5 seconds each, with the difference spectra for each segment restricted to the spectrum segment 1 … 10 Hz. The decision to use 5 second segments is based on informal empirical comparison of differently sized segments, which showed that longer segments were unduly influenced by random length pauses, while shorter segments did not capture the frequency range of interest, which roughly corresponds to syllables, words and short phrases. Four metrics were extracted from each segment: the most prominent edge frequency in the AEDS; the mean of the differences in the AES from which the AEDS is derived; the standard deviations of the spectrum segment lower than (SDleft) and higher than (SDright) the most prominent edge frequency, respectively. Additionally the overall mean line is shown, as well as a linear regression (*TaaG, Trend at a Glance*) line.

Figure 8 illustrates the variability of the most prominent edge frequency, and consequently of the associated rhythms and rhythm zone edges. Lines based on descriptive statistics derived from the rhythm zones associated with these edges, show the following patterns:

1. The values associated with the prominent frequency and with each statistic vary considerably during the utterance.

2. The NS and NNS prominent frequency variations follow a different, slower trend, with skewing of the 'sawtooth' alternations in frequency.
3. The mean over all most prominent edge frequencies is almost exactly 5 Hz, corresponding to a periodicity of 200 ms, the same for both speakers. The syllable rates in manual annotations of the same recordings are 4.16 and 3.59, respectively, both of which fall into the higher frequency zone above the most prominent edge, so possibly the edge can be interpreted as a boundary between shorter and longer units such as words or feet.
4. An interesting and more complex marker of a potential difference between the NS and NNS timing patterns is the ratio between the standard deviations of the frequency range above the most prominent edge frequency and the frequency range below this frequency: 7.4 and 2.3 respectively. A possible interpretation of these ratios is that the NNS rhythms concentrate on higher frequencies, shorter periodicities, while the NS uses varied long-term rhetorical strategies.
5. The TaaG regression line tends to fall for the NS and rise for the NNS, indicating a shift of the most prominent edge indicating a move to lower frequency edges, and the opposite for the NNS.

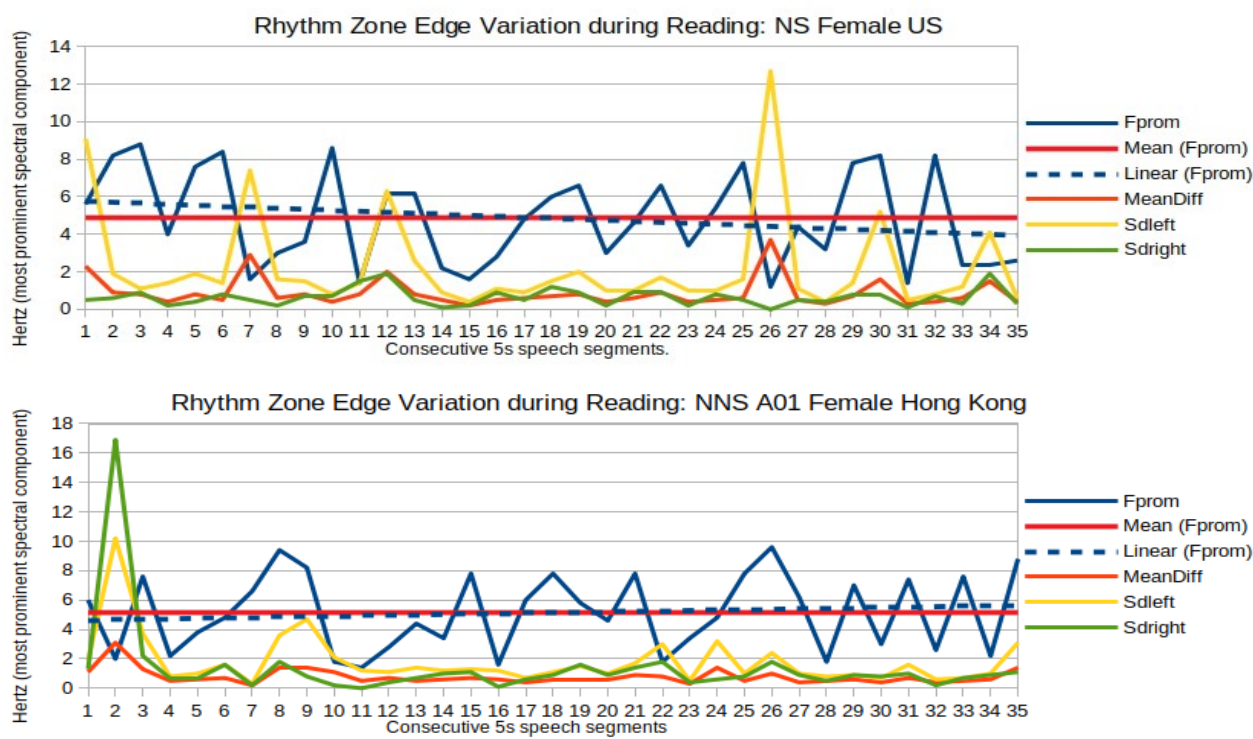

*Figure 8: Examples of English NS (US) and Cantonese NNS: the most prominent Rhythm Zone Edges in each of a series of 35 x 5s consecutive segments in the speech signal, covering 180s of the utterance.*

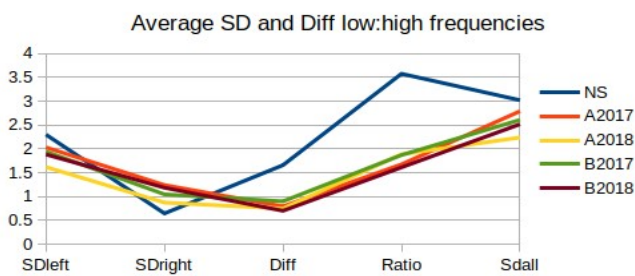

*Figure 9: Average value for frequency standard deviation markers.*

Figure 9 shows the overall values for the the rhythm zone edge markers for the NS and the two NNS readers A and B, and for the two years 2017 and 2018, revealing a strong similarity between the NNS in contrast to the considerable distance between NNS and NS. These differences suggest again that these spectral markers are not very good at distinguishing between different low performing adult non-native readings, but that the NS sub-genre can be clearly distinguished from the NNS sub-genre of reading aloud.

## 5 Summary, conclusion and outlook

The Rhythm Zone Theory (RZT) of speech rhythm was described in some detail. Unlike many previous approaches to rhythm analysis, RZT does not start with linguistic units and search for correlates, nor does it relate exclusively to expert annotation decisions on time domain concepts such as isochrony and duration variability. RZT concentrates exclusively on demodulating the physical oscillations of amplitude modulated signals in the low frequency domain, replacing the heuristic isochrony metrics with a computationally well-defined model of multiple rhythm frequencies in rhythm zones bounded by fuzzy edges in the amplitude envelope spectrum, thus extending earlier approaches to amplitude envelope spectrum analysis.

The mindset of conceptualising rhythms – in the plural – in terms of oscillations with identifiable frequencies, rather than in terms of single indices of or binary relations between the durations of speech segments, promises returns which isochrony heuristics have largely failed to find. The restriction to signal processing is, additionally, orders of magnitude faster in application and opens up a prospect of application in classification by unsupervised machine learning methods.

In an exploratory application of RZT to the analytic evaluation of fluency, it was shown that native-speaker reading can be clearly distinguished from non-native Cantonese learner readings, but that a finer classification among the non-native learners cannot be achieved with the methods used. The result gives rise to the expectation that in the long term some markers of fluency can be captured automatically and without prior annotation in easily usable online or handheld applications for direct feedback to students. In order to proceed in this direction, not only rhythm frequencies need to be captured, but also the phases of these frequencies, in order to be able to capture rhythm variations.